\documentclass[preprint]{JASA_arxiv}

\definecolor{track}{rgb}{0,0,0}
\newcommand{\modif}[1]{\textcolor{track}{#1}}

\begin{document}

\title[Noise propagation from a pair of wind turbines]{Wake‑induced variations in noise levels and amplitude modulation for two interacting wind turbines}

\author{Jules Colas}
\author{Ariane Emmanuelli}
\author{Didier Dragna}\email{didier.dragna@ec-lyon.fr}

\affiliation{Ecole Centrale de Lyon, CNRS, Universite Claude Bernard Lyon 1, INSA Lyon, LMFA,UMR5509, 69130, Ecully, France}

\author{Richard J.A.M. Stevens}

\affiliation{Physics of Fluids Group, Max Planck Center Twente for Complex Fluid Dynamics, J. M. Burgers Center for Fluid Dynamics, University of Twente, P. O. Box 217, 7500 AE Enschede, The Netherlands}

\begin{abstract}
	The influence of turbine–turbine interactions on sound propagation is investigated using numerical simulations. Three configurations are examined: turbines aligned downstream of each other, placed side by side, and arranged in a staggered pattern. The simulation framework combines large-eddy simulations for aerodynamic interactions, an aeroacoustic source model to simulate turbine sound emission, and parabolic equation methods for sound propagation. When a second turbine is positioned directly downstream, wake-induced flow focusing enhances sound pressure levels (SPL) and amplitude modulation (AM) by several decibels downwind. In side-by-side and staggered configurations, SPL increases are limited ($<2$~dBA), and AM is generally reduced due to spatial averaging. Distinct AM patterns emerge in regions where acoustic contributions from both turbines are comparable. For identical rotor speeds, AM is strongly affected by the angular offset between rotors. When rotor speeds differ slightly, beating effects occur, resulting in intermittent AM. These findings highlight the sensitivity of AM to rotor dynamics, a key factor influencing sound perception, with implications for environmental impact and turbine siting.
\end{abstract}

\maketitle

\section{Introduction}

\modif{
	The expansion of wind energy, both onshore and offshore, is a cornerstone of the global transition toward sustainable energy.
	However, noise generated by wind turbines is a critical factor influencing widespread public acceptance of wind energy.
	Studies have shown that wind turbine noise can lead to neighbor annoyance and sleep disturbance \citep{liebichSystematicReviewMetaanalysis2021, nguyenAudibilityWindFarm2022,gassnerJointAnalysisResident2022}.
	Beyond its social impact, it also carries an economical cost, as curtailment plans for noise mitigation result in loss of electrical production \citep{burkeFactorsInfluencingWind2011}.}

\modif{
	To address noise-related challenges, better layout optimization strategies that minimize noise impact while maximizing energy production are developed \citep{wuOptimizingLayoutOnshore2020,caoOptimizingWindEnergy2020}. Similarly, better curtailment plans based on accurate prediction of noise at houses are also needed \citep{nyborgOptimizationWindFarm2023}.
	Both strategies require precise modeling to ensure that noise constraints are met without unnecessarily sacrificing wind energy production.
	However, current engineering methods usually rely on simplified assumptions, such as a point source model and uniform wind speeds for propagation models.
    These methods do not reflect real-world conditions and may lead to inefficient optimization strategy.
	To overcome these limitations, high-fidelity numerical models are essential to provide detailed description of the flow around the turbines, the noise sources themselves, and the propagation of sound through the atmosphere.
}

During the past decade, significant progress has been made in these various aspects. Flow fields obtained from computational fluid dynamics simulations are now widely incorporated into propagation simulations~\cite{barlasConsistentModellingWind2017,heimannSoundPropagationWake2018}, providing more realistic representations than analytical profiles. Advanced source models, including extended and moving sources, have also been developed~\cite{tianWindTurbineNoise2016,brescianiWindFarmNoise2024,shenDevelopmentGeneralSound2022}. In parallel, several propagation models have been introduced. Wave-based approaches, based on either the parabolic equation~\cite{leePredictionFarfieldWind2016,kayserEnvironmentalParametersSensitivity2020,nyborgIntermodelComparisonParabolic2023} or the linearized Euler equations~\cite{colasImpactTwodimensionalSteep2024} are the state of the art for wind turbine noise \modif{propagation} studies. While they have been restricted to two-dimensional configurations, first simulations using three-dimensional parabolic equations have been recently reported~\cite{BOMMIDALA2025119036}, although limited to low frequencies.

\modif{The literature on wake-induced propagation effects has primarily focused on isolated wind turbines, with limited research on noise propagation from multiple turbines.}
So far,  the effects of successive turbine wakes for a row of aligned wind turbines have been examined by \citet{sunDevelopmentEfficientNumerical2018} using a single point source at the hub.  \citet{shenAdvancedFlowNoise2019} have illustrated the overall sound pressure level for a wind farm of 25 turbines in real complex terrain.
\modif{
Simulations of the complete wind farm, accounting for both wake interactions and flow propagation effects, were also performed within an optimization framework \citet{nyborgOptimizationWindFarm2023}.
}
\modif{
	Despite these studies, the effect of multiple wakes on sound propagation  has not been systematically examined.}
\modif{
	Moreover, propagation effects on amplitude modulation (AM), a critical aspect of wind turbine noise perception and regulation \cite{leeAnnoyanceCausedAmplitude2011, hansenPrevalenceWindFarm2019}, has not been thoroughly explored in the context of multiple turbines.
}

In this study, we aim to address these gaps by considering noise propagation from a pair of wind turbines as a canonical case for multiple wind turbines.
\modif{
	Specifically, we investigate the effects of turbine-turbine interactions and multiple wakes on noise generation and propagation with an extended focus on AM.
}
Four configurations are considered: a single isolated turbine as a reference, and a pair of turbines arranged either side by side, one behind the other, or in a staggered layout.
\modif{
	Focusing on two turbines helps isolate propagation effects that may occur in larger wind farms.
	These four configurations reflect possible relative turbine arrangement within a larger wind farm.
}



The paper is organized as follows. In Sec.~\ref{methods.sec}, the four configurations and the numerical methods are introduced. Sec.~\ref{inter_results.sec} presents the mean flow, the source characteristics, and the propagation effects. Sec.~\ref{noise.sec} details the average sound pressure level and the amplitude modulation around the wind turbine pair. Finally, concluding remarks are given in Sec.~\ref{conclu.sec}.

\section{Scenario and methods} \label{methods.sec}

\subsection{Configurations}
\begin{figure*}[h!tb]
	\includegraphics{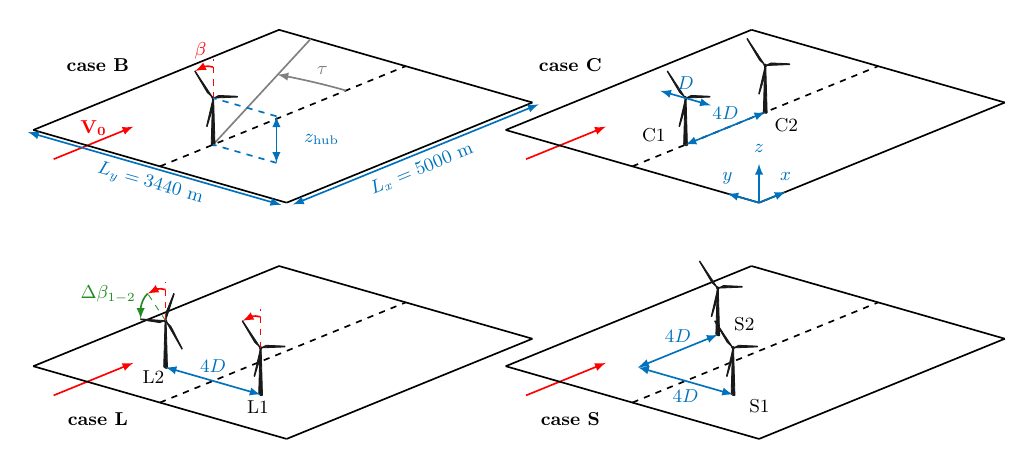}
	\caption{Sketch of the four configurations.}
	\label{f5: case_def}
\end{figure*}

Four configurations are considered, as illustrated in Fig.~\ref{f5: case_def}. First, a baseline case with a single turbine is simulated (case B). It is then compared with three layouts of wind turbine pairs: column (case C), line (case L) and staggered (case S) layouts. The turbines are spaced out by four rotor diameters (480~m) in the $x$- and $y$-directions.
\modif{In the following, indices 1 and 2 refer to the upstream and downstream turbines in cases C and S, respectively, and to the left and right turbines (relative to the flow direction) in case L, as shown in Fig.~\ref{f5: case_def}.}

The turbines are all identical and have a hub height $z_{\rm hub}=90~$m. The geometry of the blades is that proposed in~\citet{tianWindTurbineNoise2016}, except that the rotor diameter is scaled up to $D=120~$m so that the dimensions of the wind turbines correspond to those of the 5~MW NREL wind turbine \citep{jonkmanDefinition5MWReference2009}.

A stable atmospheric boundary layer (ABL) is considered because it is generally considered the worst-case scenario for wind turbine sound propagation~\citep{barlasVariabilityWindTurbine2018,heimann3DsimulationSoundPropagation2018}. The stable atmosphere is simulated with a cooling rate equal to -0.2~K~h$^{-1}$, a roughness length equal to 0.1~m and a friction velocity of 0.6~m~s$^{-1}$ to obtain a wind speed at hub height equal to 11.4~m~s$^{-1}$.

\subsection{Methods}

Wind turbine noise propagation requires three elements: a description of the atmospheric flow, a model for the wind turbine source, and a model for sound propagation.

\subsubsection{Atmospheric flow}
The atmospheric flow is obtained using large-eddy simulations (LES). \modif{A key benefit of LES is its ability to accurately capture turbulent exchange processes in the atmosphere, particularly within wind farms where multiple wakes interact \cite{blockenRANSBuildingSimulation2018a}. The code solves the incompressible Navier-Stokes equations with Boussinesq approximation.
	A Lagrangian-averaging scale-dependent model \cite{gaddeLargeEddySimulationsStratified2021} is used to obtain an accurate representation of the sub grid scale properties.
}
\modif{The domain is discretized with a grid step of 15~m, 15.6~m and 10.4~m in the $x$-, $y$- and $z$- directions respectively.
	Periodic boundary conditions are used in the $x$- and $y$- directions, a free-slip boundary condition is used at the top of the domain and Monin Obukhov similarity theory is applied to model wall shear stress and buoyancy flux at the ground \cite{beareIntercomparisonLargeEddySimulations2006}.
	A concurrent precursor method~\cite{stevensConcurrentPrecursorInflow2014} is used
	meaning that two LES are run concurrently, the first one, without turbines, provides a realistic ABL that is used as input for the second simulation where the wind turbines are placed inside the flow.
}
An actuator disk approach is used to model the wind turbine~\cite{stevensComparisonWindFarm2018}. The LES is run until a statistical stationary solution is obtained. 
\modif{Although the LES provides time-dependent results, only the time-averaged flow fields are used in this work. They are determined by time averaging the instantaneous fields over four hours.
	For simplicity, we use a constant temperature and sound speed in the following.}

\modif{	In an acoustic application, \citet{barlasEffectsWindTurbine2017} compared different flow inputs for an isolated wind turbine and showed that using LES time-averaged flow field inputs gave similar propagation results to fully unsteady simulations with time-dependent flow inputs. However, unsteady effects may be relevant in multi-turbine scenarios and should be investigated in future work. Nevertheless, considering atmospheric turbulence significantly increases the computational cost of the acoustic simulations, as multiple propagation simulations must be performed for several snapshots of the flow field. 	Furthermore,  the turbulence length scales resolved by LES are too large to correctly capture noise scattering in the frequency range of interest \cite{barlasEffectsWindTurbine2017}. Thus, a dedicated method is required to recover information from unresolved turbulence scales.
}

\subsubsection{Source model}

The source model is that developed by~\citet{tianWindTurbineNoise2016}. Each blade is split into segments with a sufficiently large span compared to its cord \modif{(an aspect ratio of 3 is used in this work)}, along which the geometrical and flow properties are assumed to be constant. Amiet's theory \citep{amietAcousticRadiationAirfoil1975,amietNoiseDueTurbulent1976} is then applied to determine the turbulent inflow noise (TIN) and the trailing edge noise (TEN) due to each segment.

\modif{
	The LES data in the rotor plane is used to specify the mean flow velocity and the turbulent dissipation rate, which are required for TIN and TEN computations.}
\modif{
	The spectrum of the velocity fluctuations in the incoming flow is modeled using the Kolmogorov spectrum as in~\citet{mascarenhasSynthesisWindTurbine2022}, from which the TIN is derived.}
\modif{
	The effective wind speed for each blade segment is computed based on the wind speed obtained from  the LES and the rotational speed of the blade.
	The boundary layer properties at the trailing edge are then determined using Xfoil~\cite{drelaXFOILAnalysisDesign1989}.
	They are subsequently used to  model the wall pressure fluctuation spectrum \cite{leePredictionAirfoilTrailingEdge2019} required to compute the TEN. }
Finally, the source model provides the sound pressure level in the free field, denoted ${\rm SPL}_{\rm ff}(m,\beta,f)$, for each blade segment indexed by $m$ as a function of the frequency $f$ and the rotor angle $\beta$.

As the flow in the rotor plane is different for each turbine, it is also necessary to specify a control law that relates the rotational speed of the rotor $\Omega$ to the impinging flow.
\modif{
	For that, we use the relation that gives $\Omega$ from the wind speed at the hub $u_{\rm hub}$ for the NREL 5MW wind turbine to compute a constant $\Omega$ for each turbine in each configuration.}
The corresponding values in the cut-in, rated and cut-out conditions are indicated in Table~\ref{t2: rotor}.
\modif{Note that $\Omega$ does not increase linearly between the cut-in and rated wind speeds (see Fig.~9-1 in \citet{jonkmanDefinition5MWReference2009}).}

\begin{table}[htbp]
	\centering
	\caption{Wind turbine control properties from \citet{jonkmanDefinition5MWReference2009}.}\label{t2: rotor}
	\begin{tabular}{lccc}
		\hline
		\hline
		                      & cut-in       & rated           & cut-out       \\
		\hline
		wind speed at the hub & 3 m~s$^{-1}$ & 11.4 m~s$^{-1}$ & 25 m~s$^{-1}$ \\
		rotor speed           & 6.9 rpm      & 12.1 rpm        & 0 rpm         \\
		\hline
		\hline
	\end{tabular}
\end{table}

\subsubsection{Propagation model}

Sound propagation in the atmosphere is determined by solving a parabolic equation (PE) proposed in~\citet{ostashevWaveExtrawideangleParabolic2020} that accounts for a moving atmosphere without relying on an effective sound speed approximation. The results in~\citet{colasWindTurbineSound2023a} have shown that this PE yields excellent agreement with numerical solutions of the linearized Euler equations for a wind turbine in flat terrain. Note that a modified PE that better handles the sound speed gradients has also been proposed and analyzed in \citet{ ostashevPhasepreservingNarrowWideangle2024}.

The propagation simulation is carried out using the N$\times$2D-PE propagation method: two-dimensional simulations are performed in vertical planes for several azimuthal angles $\tau$  around each source. This approach provides three-dimensional acoustic fields but neglects horizontal refraction. The PE is solved using the Crank-Nicolson method with a uniform grid step set to a tenth of the wavelength. A perfectly matched layer is used as a non-reflecting boundary condition at the top of the domain. At the ground, an impedance boundary condition is implemented. The surface impedance is prescribed using the variable porosity model \cite{attenboroughOutdoorGroundImpedance2011} with a flow resistivity of 50~kN~s~m\textsuperscript{-4} and a porosity change rate of 100 m\textsuperscript{–1}, representing a grassy ground. In addition, the projection of the horizontal mean flow velocity in the propagation plane is determined from the LES data. The sound speed is constant and is set to $c_0=340$~m~s\textsuperscript{-1}.

The rotor is discretized with an angular step $\Delta \beta = 10^{\circ}$ and each blade is divided into 8 segments. This  yields a total of 288 source positions for each turbine.  To reduce the number of simulations and the computational cost, the approach proposed in~\citet{cotteExtendedSourceModels2019} and used in~\citet{colasWindTurbineSound2023a} is employed:  propagation simulations are only performed for seven fictive sources distributed at different heights along the rotor centreline. The sound pressure level relative to the free-field level for each actual source position, denoted $\Delta L(m,\beta, f)$, is then determined from those of the fictive sources using interpolation.
\modif{Comprehensive definition of $\Delta L$ (and other quantities such as SPL$_{\rm ff}$) and further details on their computations can be found in \citet{colasWindTurbineSound2023a} and \citet{cotteExtendedSourceModels2019}.}

The computational domain around each wind turbine extends 1~km upstream, 4~km downwind and 3.44~km crosswind. An angular step of 2$^{\circ}$ is used to discretize the whole domain except in the downwind direction ($-16^{\circ}<\tau<16^{\circ}$) where a 1$^{\circ}$ angular step is used. This is done to better account for the wake effect in this direction. The results are then interpolated onto a uniform Cartesian grid with a grid step of 10~m to facilitate post-processing in configurations involving multiple wind turbines.
\modif{
	The simulations are performed for a set of frequencies between 50~Hz and 1080~Hz (same as \citet{colasWindTurbineSound2023a}) that were tested to obtain converged values of overall SPL and AM for each third-octave frequency band \citep{cotteCouplingAeroacousticModel2018}.
}

\subsubsection{Chaining methodology}

The chaining methodology detailed in~\citet{colasWindTurbineSound2023a} is employed.
The flow properties are used to feed the source and propagation models.
The sound pressure levels (SPL) from the wind turbines are obtained by combining the free-field levels predicted by the source model with the relative levels determined by the propagation model.

In more details, the SPL due to a given segment is computed with:
\begin{equation}
	{\rm SPL}(m,\beta,f) = {\rm SPL}_{\rm ff}(m,\beta,f) + \Delta L(m,\beta,f) - \alpha(f) R
\end{equation}
with $\alpha$ the atmospheric absorption coefficient computed from~\citet{isoAcousticsAttenuationSound1993} and $R$ the source-receiver distance. To compute the SPL generated by two turbines operating at different rotor speeds, it is more convenient to work with time-domain signals rather than angular-domain signals. To do so, we express the SPL as a function of the receiver time, as done in~\citet{mascarenhasSynthesisWindTurbine2022}. We first note that the time-domain signal is periodic with period $2\pi/(3\Omega)$.  Considering a time step $\Delta t$, the SPL due to a blade located at $t=0$ at $\beta=0^{\circ}$ is written:
\begin{equation}
\modif{	\text{SPL}(t_k,f) = 10\log_{10}\left(\sum_m 10^{\text{SPL}(m,\beta_j,f)/10}\,\delta_{m,\,j,\,k}\right),}
\end{equation}
with $t_k=k\Delta t$. The function $\delta_{m,\,j,\,k}$ is given by:
\begin{equation}
	\delta_{m,\,j,\,k} =
	\begin{cases}
		1, \quad \text{if} \quad T_{m,\,j}\leq t_k<T_{m,\,j+1} \\
		0, \quad \text{otherwise}
	\end{cases}
\end{equation}
where $T_{m,\,j}$ is the arrival time at the receiver of the contribution from the segment $m$ at rotor angle $\beta_j = j\Delta \beta$. For simplicity, we assume that the propagation occurs along a straight line with constant sound speed $c_{\rm ref}=c_0$, leading to:
\begin{equation}
	T_{m,\,j}  = \dfrac{R_{m,\,j}}{c_{\rm ref}}+\dfrac{\beta_j}{\Omega}
\end{equation}
where $R_{m,\,j}$ is the distance from the blade segment to the receiver.

The SPL due to the wind turbine as a function of time is then obtained by logarithmically summing the time signals of the three blades. The same methodology is applied to compute the SPL due to two turbines. Finally, the \modif{overall SPL (OASPL)} is obtained by integrating the SPL over the frequency band of interest. Two quantities of interest are derived from the OASPL: the average OASPL, denoted $\overline{\rm OASPL}$, and the amplitude modulation (AM), defined as the difference between the maximum and minimum values of the OASPL.

The time step $\Delta t$ must be chosen smaller than the minimum value of $\Delta \beta/\Omega$, which is approximately 0.14~s.  In the following, $\Delta t$ is set to 0.1~s. We have checked that further reduction of $\Delta t$ yields negligible changes in the average OASPL and AM.

\section{Flow, source, and propagation} \label{inter_results.sec}

\subsection{Flow}\label{s: flow}
\begin{figure*}[h!tb]
	\centering
	\includegraphics{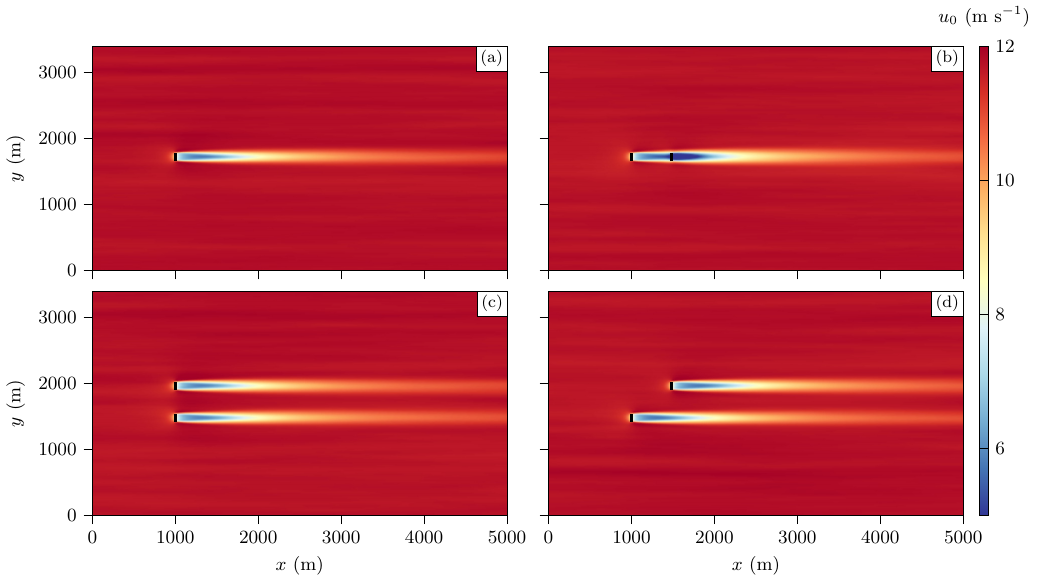}
	\caption{Magnitude of the mean flow streamwise velocity at hub height, \modif{computed with LES}, for the four cases: (a) B, (b) C, (c) L, and (d) S. The rotors of the wind turbines are represented by a thick black line.}
	\label{f5: cases_u}
\end{figure*}

The mean streamwise velocity $u_0$ is shown at hub height for the four configurations in Fig.~\ref{f5: cases_u}. For case B  in Fig.~\ref{f5: cases_u}~(a), we observe the wake behind the turbine, which extends downstream up to several kilometers. For case C in Fig.~\ref{f5: cases_u}~(b), we note the two wakes superimposing on each other, creating a stronger velocity deficit behind the second turbine. For cases L (Fig.~\ref{f5: cases_u}~(c)) and S (Fig.~\ref{f5: cases_u}~(d)), the wakes behind each turbine can be distinguished, without any noticeable wake interaction.

\modif{
    Small spanwise variations are also visible for all the cases.
 These small wind-speed variations are a result of limited time averaging in our spanwise-periodic simulations; in the limit of very long averaging, these spanwise variations would ultimately average out. 
 This can take a very long time, as documented in the literature \citep{stevensConcurrentPrecursorInflow2014,muntersTurbulentInflowPrecursor2016}. 
 In the real atmosphere, incoming wind speed also varies across a wind farm due to large-scale atmospheric flow features. 
 Given that small variations in incoming wind speed between turbines, even after averaging over hours, are realistic, we choose to keep these spanwise variations.}

\subsection{Source} \label{s5: source}

The source model uses as input the profiles of the mean flow velocity and of the turbulent dissipation rate in the rotor plane of the turbines. These profiles are all nearly identical, except for turbine C2 which is positioned in the wake of turbine C1. The source model provides as output the SPL in the free field for each blade segment at every angular position. From this, we can determine the sound power level in the free field, denoted SWL, and the overall sound power level integrated over the frequency band of interest, denoted OASWL. As for the sound pressure level, averaged values of the power levels and associated AM can then be computed.

\begin{figure}[h!tbp]
	\centering
	\includegraphics{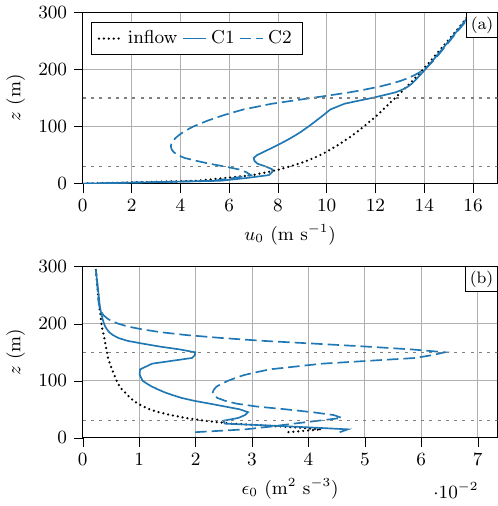}
	\caption{Profile of (a)~the streamwise velocity $u_0$ and (b)~the turbulent dissipation rate $\epsilon_0$ along the rotor centreline for the two wind turbines in case C. The rotor edges are visualized with gray dotted horizontal lines. The inflow profiles are also represented for reference.}
	\label{f5: SWL_compar_1D_2D_flow}
\end{figure}

Note that the sound powers presented below do not correspond to the actual ones, determined from the integration of the acoustic field over a surface enclosing the source. Instead, they are effective powers obtained from a single receiver located downwind. In more detail, the power levels are computed from the relation ${\rm SWL} =  {\rm SPL}_{\rm ff} + 10\log_{10}(4\pi R^2)$, where the sound pressure level in the free field ${\rm SPL}_{\rm ff}$ is evaluated at a downwind receiver sufficiently far from the turbine
\modif{(3~km downwind)}
and $R$ is the source-receiver distance. This effective power is introduced to compare sound production by the turbines in the downwind direction.
Figure~\ref{f5: SWL_compar_1D_2D_flow} shows the profiles of the mean streamwise velocity and the turbulent dissipation rate $\epsilon_0$ along the rotor centreline for both turbines in case C. The corresponding profiles of the ABL flow are also plotted as reference.  We note for C1 a reduction of the wind speed at the rotor due to a blockage effect and the increase of the turbulence dissipation rate at the outer part of the rotor. For C2, as it is positioned in the wake of C1, the profiles are significantly different. Along the rotor, the velocity decreases compared to that of C1. Simultaneously, the turbulent dissipation rate strongly increases, especially at the top of the rotor.

\begin{figure}[htbp]
	\centering
	\includegraphics{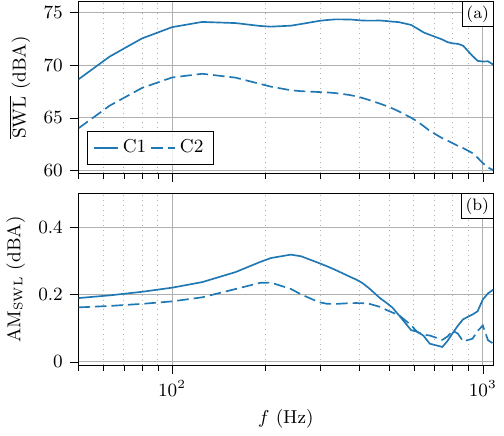}
	\caption{Spectrum of (a) the average sound power level $\overline{\rm SWL}$ and (b) the corresponding amplitude modulation AM$_{\rm SWL}$ for the two turbines in case C.}
	\label{f5: SWL_compar_1D_2D_omega_C2}
\end{figure}

The effects of the different flow profiles on the source model are illustrated in Fig.~\ref{f5: SWL_compar_1D_2D_omega_C2}. Fig.~\ref{f5: SWL_compar_1D_2D_omega_C2}~(a) shows the spectrum of the average source power level in the free field $\overline{\rm SWL}$ for the wind turbines C1 and C2. The spectrum for C1 is composed of two humps of equal amplitude, the low-frequency one corresponding to TIN and the high-frequency one to TEN. The spectrum for C2 shows a global reduction in its amplitude because of the smaller wind velocity in the rotor plane. The hump at low frequencies has a higher amplitude than that at high frequencies, because of the relative increase in TIN due to the increase in turbulent dissipation rate. The spectrum of amplitude modulation due to the source AM$_{\rm SWL}$ is plotted in Fig.~\ref{f5: SWL_compar_1D_2D_omega_C2}~(b). It is small for both wind turbines, with values below 0.3~dBA. Note that the large spatial variations in the profiles for C2 do not lead to a significant increase in AM associated to the source. As a reminder, these source properties are calculated from a receiver located downwind. In particular, amplitude modulation is expected to be stronger in the crosswind direction.

\begin{table}[ht!]
	\centering
	\caption{Source parameters computed for each wind turbine.}\label{t5: source}
	\modif{
		\begin{tabular}{lccccccc}
			\hline
			\hline
			                                        & B     & C1    & C2    & L1    & L2    & S1    & S2    \\
			\hline
			$u_{\rm hub}$ (m s$^{-1}$)              & 9.00  & 8.92  & 4.10  & 8.98  & 9.02  & 8.92  & 9.00  \\
			$\epsilon_{\rm hub}$ (m$^{2}$ s$^{-3}$) & 0.012 & 0.012 & 0.024 & 0.012 & 0.012 & 0.012 & 0.012 \\
			$\Omega$ (rpm)                          & 10.3  & 10.2  & 7.1   & 10.3  & 10.3  & 10.2  & 10.3  \\
			$\overline{\rm OASWL}$ (dBA)            & 103.4 & 103.2 & 95.8  & 103.4 & 103.5 & 103.2 & 103.5 \\
			\hline
			\hline
		\end{tabular}
	}
\end{table}

\begin{figure*}[h!tbp]
	\centering
	\includegraphics{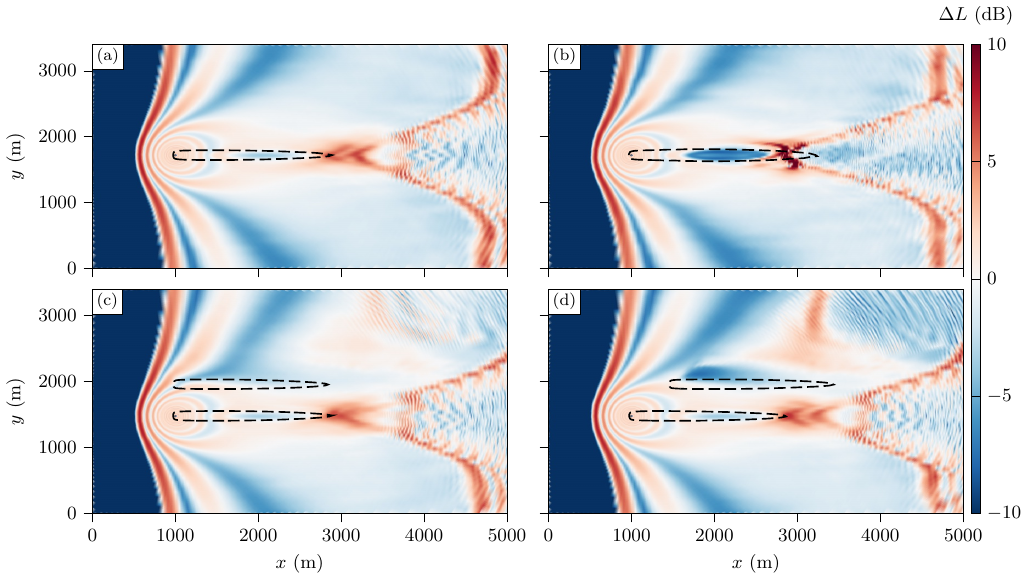}
	\caption{Sound pressure level relative to the free-field level $\Delta L$ averaged over the third-octave band centered at 1~kHz at $z=2~$m for a source at the hub height for the wind turbines: (a) B, (b) C1, (c) L1, and (d) S1. The black dashed lines correspond to $u_0(z=90 \mathrm{m})=10$~m s$^{-1}$ and aim at highlighting the wake regions.}
	\label{f5: dl_top}
\end{figure*}

For the four configurations, Table~\ref{t5: source} gives  the mean flow velocity and the turbulent dissipation rate at the hub, denoted $u_{\rm hub}$ and $\epsilon_{\rm hub}$,  as well as the rotational speed of the rotor and the average OASWL of the turbines. These values are almost the same for all turbines except for C2. Indeed, the mean flow velocity at the hub for C2 is significantly lower, resulting in a lower rotational speed.
\modif{Consequently}, the sound power level of C2 is reduced by more than 7~dBA.
The wind speed at the hub differs by less than 0.5~\% for L1 and L2, leading to
identical rotor speeds (with a controler precision of 0.1 rpm).
On the other hand, for S1 and S2, the deviations in wind speed at hub height are slightly larger, around 0.9~\% leading to different rotor speeds.
\modif{ These variations are caused by the spanwise wind speed differences discussed in Sec.~\ref{s: flow}.
Although minor, these differences in rotor speed affect the temporal signals, creating a beating effect (explained in Sec.~\ref{synchr.sec})}

\subsection{Propagation} \label{s5: deltaL}

This section presents the effect of the mean flow on sound propagation. For clarity, the analysis is limited to a single source positioned at the turbine hubs.

\begin{figure}[htbp!]
	\centering
	\includegraphics{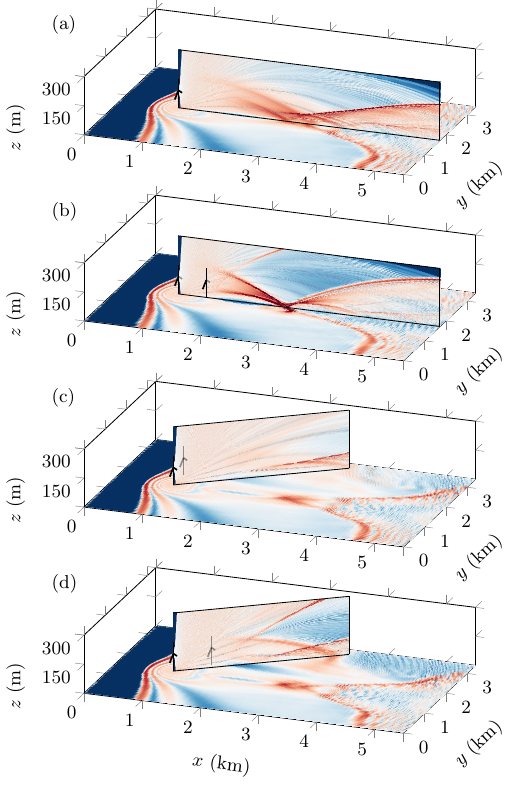}
	\caption{Sound pressure level relative to the free-field level $\Delta L$ averaged over the third-octave band centered at 1~kHz for a horizontal plane at 2~m height and a vertical plane at (a, b)~$\tau=0$°  and (c, d) $\tau=35$°. The source is located at the hub of the wind turbines: (a)~B, (b)~C1, (c)~L1, and (d)~S1. The color range is the same as in Fig.~\ref{f5: dl_top}.}
	\label{f5: dl_3D}
\end{figure}

Fig.~\ref{f5: dl_top} shows the $\Delta L $ at 2~m height averaged over the third-octave band centered at 1~kHz for a point source at the hub of turbines B, C1, L1, and S1. The contour lines of the mean flow streamwise velocity  for 10~m s$^{-1}$ are overlaid as black dashed lines to indicate the wake regions. Fig.~\ref{f5: dl_3D} displays the same data in a  3D view to highlight the influence of multiple wakes on noise amplification at the ground.
For case B in Fig~\ref{f5: dl_top}~(a) and Fig~\ref{f5: dl_3D}~(a) we note several characteristics of sound propagation in the flow of an isolated wind turbine~\cite{barlasEffectsWindTurbine2017,heimannSoundPropagationWake2018,colasImpactTwodimensionalSteep2024}. Upwind, a shadow zone is generated for $x<800~$m with high levels at its boundary. Downwind, an amplification of the sound levels at the ground is seen in the flow direction ($\tau =0^{\circ}$) at $x=3000~$m due to wake focusing. As the propagation angle increases, focusing at the ground moves away, resulting in a V-shaped amplification zone. Finally, for sufficiently large propagation angles, sound waves do not cross the wake. Focusing observed at $x=4700~$m, independent of the $y$-direction, is instead caused by the velocity gradient in the ABL.

The wake superposition in case C has a significant effect on sound propagation downwind, see Fig.~\ref{f5: dl_top}~(b). Focusing in the flow direction is stronger and closer to the turbine by 100~m compared to the baseline case. This is induced by the second wake as shown in Fig~\ref{f5: dl_3D}~(b). Two focusing zones appear just after the second turbine due to the large velocity gradients at the top and bottom of the rotor that efficiently redirect sound waves toward the ground. This is similar to what was shown in \citet{sunDevelopmentEfficientNumerical2018}.
The V-shaped pattern in the $\Delta L$ field after this initial focusing is still visible and does not seem to be strongly affected by the second wake.

The second wake also has an effect for layouts L and S, but it is less pronounced. Figs.~\ref{f5: dl_top}~(c) and \ref{f5: dl_top}~(d) show an increase in $\Delta L$  downstream for propagation directions that intersect the wake of the second turbine. This effect is more pronounced in case S than in case L. In fact, the impact of the wake depends not only on the velocity deficit but also on the propagation angle. Specifically, for a propagation angle $\tau$, considering only the streamwise component of the mean flow velocity, the wind speed projected onto the propagation direction is $u_0 \cos\tau$. For a given velocity deficit, the influence of the second wake becomes more significant as $\tau$ approaches zero. Consequently, in case S, as propagation angles intersecting the second wake are smaller, refraction is stronger. Conversely, in case L, the propagation angles intersecting the second wake are larger, leading to small variations in the projected wind speed and, therefore, to reduced refraction effects by the second wake. This is illustrated in Figs.~\ref{f5: dl_3D}~(c) and \ref{f5: dl_3D}~(d) where focusing induced by the second wake is more noticeable in case S than in case L for the same propagation angle ($\tau=35$°). In addition, focusing caused by the ABL velocity gradient at $x=4700~$m visible for cases B and C in Figs.~\ref{f5: dl_top}~(a) and (b) is no longer observed for $y>2500$~m for cases L and S in Figs.~\ref{f5: dl_top}~(c) and (d). Finally, a region of lower $\Delta L$ is also present just after the second wake in Fig.~\ref{f5: dl_3D}~(d).

\begin{figure}[htbp!]
	\centering
	\includegraphics{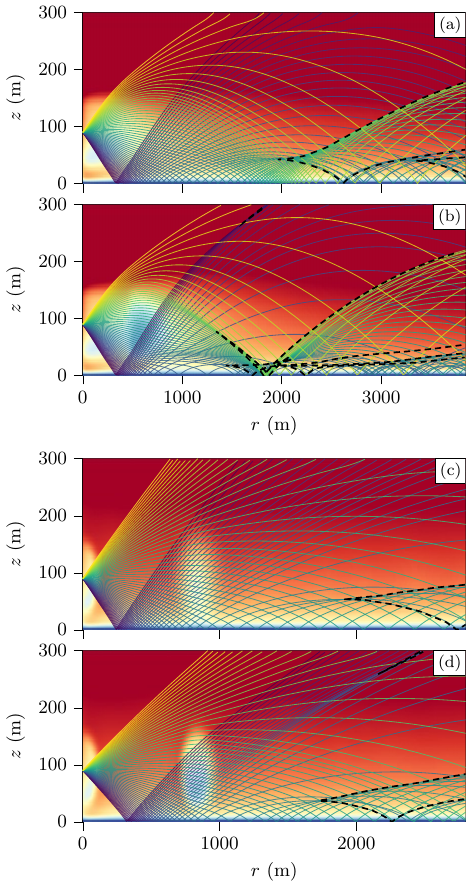}
	\caption{Ray tracing for turbine~(a) B and (b)~C1 for a propagation angle $\tau=0^{\circ}$ and for turbine (c)~L1  and (d)~S1 for $\tau=35^{\circ}$. Rays are superimposed on the projected horizontal flow velocity field in the propagation plane, while caustics are depicted as black dashed lines. The color range for the velocity magnitude is the same as in Fig.~\ref{f5: cases_u}.}
	\label{f5: rays_1}
\end{figure}

To help understand the refraction of sound waves by the flow, ray-tracing simulations are carried out following \citet{candelNumericalSolutionConservation1977} and \citet{scottWeaklyNonlinearPropagation2017}.  The rays are presented in Fig.~\ref{f5: rays_1} for side planes corresponding to those shown in Fig.~\ref{f5: dl_3D}. They are superimposed  onto the projection of the horizontal component of the flow velocity onto the considered plane. In Fig.~\ref{f5: rays_1}, $r$ denotes the horizontal distance from the source.  For the baseline case B in Fig~\ref{f5: rays_1}~(a), we observe focusing induced by the wake. This leads to a high density of rays close to the ground for 1900~m$<r<2000~$m, which corresponds to the increase in $\Delta L$ at the ground in Fig.~\ref{f5: dl_top}~(a). Moreover, a caustic starting at $r=2000$~m and $z=50~$m is also observed, and its position coincides with the rise in $\Delta L$  visible  in Fig.~\ref{f5: dl_3D}~(a).

For turbine C1 (Fig.~\ref{f5: rays_1}~(b)),  the wake of turbine C2 acts as a converging lens that enhances focusing. This leads to the formation of a caustic in the wake of C2, with a focal point at $r=1340$~m and $z= 78$~m, which  impacts the ground around $r=1900$~m. This explains the large increase in $\Delta L$ observed at this position in Fig.~\ref{f5: dl_3D}~(b). 

The difference between the refraction induced by the second wake for cases L and S is illustrated in Figs.~\ref{f5: rays_1}~(c) and (d) for $\tau = 35$°. Since the velocity deficit due to the wake in the propagation plane is more pronounced in case S than in L, the refraction induced by the second wake is stronger in the former case.

\begin{figure*}[htbp]
	\centering
	\includegraphics{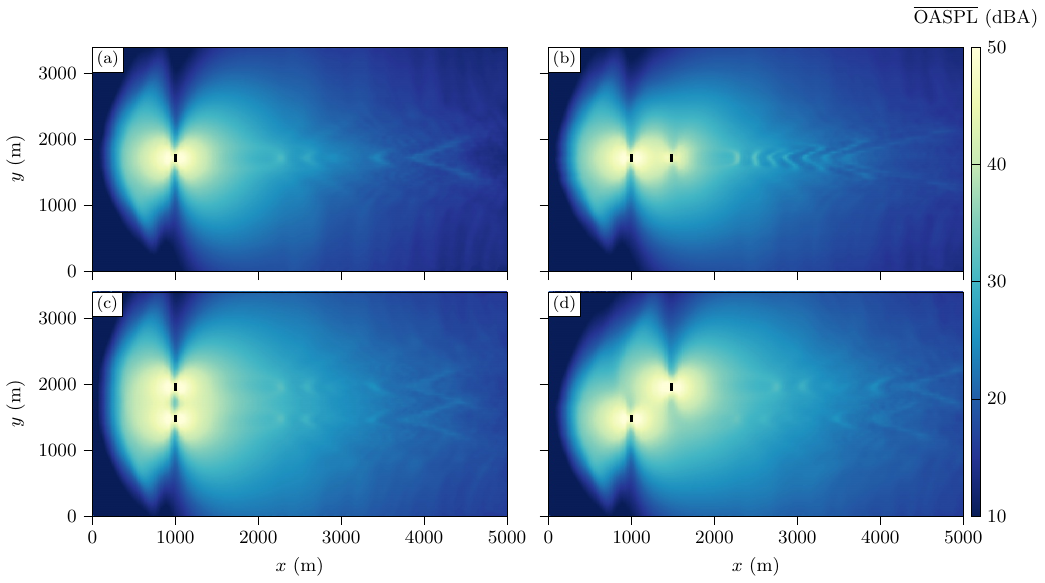}
	\caption{Average overall sound pressure levels $\overline{\rm OASPL}$ at $z=2~$m for the four cases: (a) B, (b) C, (c) L, and (d) S.}
	\label{f5: oaspl_top}
\end{figure*}

The converging lens effect induced by the second wake is clearly visible in case S in Fig.~\ref{f5: rays_1}~(d). Bundles of rays passing through the second wake are refracted inward, resulting in amplified sound pressure levels. Thus, the rays launched downwards and then reflected from the ground before crossing the second wake generate a caustic at $r=2150$~m and $z=260$~m in Fig.~\ref{f5: rays_1}~(d). Similarly, the region of increased levels visible in Fig.~\ref{f5: dl_3D}~(d) that intersects the ground at a large distance can be associated with a bundle of rays launched upwards, then refracted downwards by the ABL velocity gradient, and finally focused due to the crossing of the second wake. We also observe that the presence of the second wake tends to shift the caustic associated with the ABL velocity gradient closer to the source: thus, this caustic is located at the ground level 4.3~km from the source without a second wake, but 2.9~km in case L and 2.3~km in case S. Finally,  the coarser ray density after the second wake for case S in Fig.~\ref{f5: rays_1}~(d) is in agreement with the reduced value of $\Delta L$  in Fig.~\ref{f5: dl_3D}~(d).

\section{Noise from the wind turbine pair} \label{noise.sec}

Using the methodology from Sec.~\ref{methods.sec},  the instantaneous OASPL fields are computed from the results in Sec.~\ref{inter_results.sec}, accounting for the rotation  of the wind turbine blades. From these, the average OASPL and the AM fields are determined.

\subsection{\modif{Overall sound pressure level}}\label{oaspl.sec}

\begin{figure}[h!tbp]
	\centering
	\includegraphics{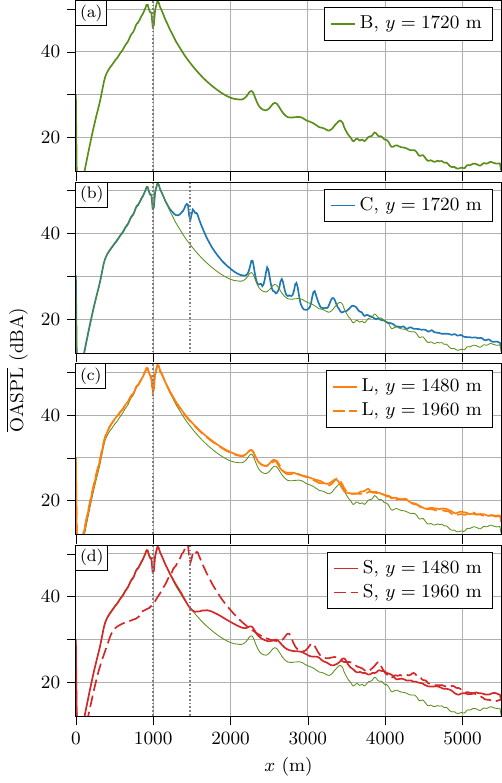}
	\caption{Average overall sound pressure levels $\rm \overline{OASPL}$  at $z=2~$m along lines of receivers whose $y$-coordinate corresponds to the turbine positions in the four cases: (a) B, (b) C, (c) L, and (d) S. 
		The gray dotted vertical lines correspond to the wind turbine positions. For comparison purposes, the result for B in (a) is reproduced in (b-d).}
	\label{f5: oaspl_compar_line}
\end{figure}

The $\rm \overline{OASPL}$ fields at 2~m height are shown in Fig.~\ref{f5: oaspl_top} for the four cases. For the baseline case B in Fig.~\ref{f5: oaspl_top}~(a), we find several characteristics of the noise generated by an isolated wind turbine, including  a dipolar directivity with cross-wind extinction zone, a shadow zone upwind, for $x<400~$m, and focusing patterns downwind due to the wake. Fig.~\ref{f5: oaspl_top}~(b) shows the $\rm \overline{OASPL}$ field for case C with the contributions of the two wind turbines accounted for. The downstream turbine (C2) contribution is smaller due to the modified inflow conditions induced by the upstream turbine wake causing a reduction in sound power level (as described in Sec.~\ref{s5: source}). The upwind shadow zone and crosswind extinction zones of the first turbine closely resemble those of case B. The crosswind extinction zones of the second turbine are barely visible due to the contribution of the first turbine that brings significant energy in these zones. The downwind focusing pattern also shows some differences, with more pronounced focusing in this case. The amplification at the ground is primarily due to the first wind turbine emission, which is strongly refracted by the second turbine wake (as shown in Sec.~\ref{s5: deltaL}). 

In Figs.~\ref{f5: oaspl_top}~(c) and (d), cases L and S display downwind focusing patterns that appear very similar to those obtained from a single wind turbine, in accordance with the limited interaction between the wakes observed for the mean velocity fields in Fig.~\ref{f5: cases_u}.
\modif{
    Note that, in rural environments, background noise is typically around 30 dBA.
	Therefore, noise generated by a wind turbine below this level would likely go unnoticed.
	Nonetheless, the analysis of propagation effects and the comparison between cases is performed even for OASPL below this value.
    It provides fundamental insights into sound propagation effects and the impact of relative turbine positioning, with findings that could apply to larger wind turbines or varying atmospheric conditions where turbine noise levels might be higher.}

The $\rm \overline{OASPL}$ for the different cases is plotted in Fig.~\ref{f5: oaspl_compar_line} along lines of receivers whose $y$-coordinate corresponds to the turbine positions at a height of 2~m. For case B in Fig.~\ref{f5: oaspl_compar_line}~(a), we note the overall decrease of the $\rm \overline{OASPL}$ with distance, with a sharp decrease upwind for $x<400$~m due to the shadow zone. Wake focusing causes bumps in $\rm \overline{OASPL}$ downwind at $x=2250$~m and 3400~m, with a maximum increase of 3~dBA.

Cases C, L, and S in Figs.~\ref{f5: oaspl_compar_line}~(b), (c), and (d), respectively, show an overall increase downwind compared to the baseline case B due to the presence of two wind turbines instead of one. In particular, for case C in Fig.~\ref{f5: oaspl_compar_line}~(b), the $\rm \overline{OASPL}$ rises significantly around $x= 1480~$m due the second wind turbine.
Again for case C, focusing attributed to the superposition of the two wakes is visible for 2000~m$<x<$3500~m with several narrow peaks. Additional simulations, not shown here for brevity, indicate that increasing the number of fictive sources in the source model introduces more peaks in the $\rm \overline{OASPL}$ and fills the gaps between those in Fig.~\ref{f5: oaspl_compar_line}~(b). As a consequence, a continuous increase in $\rm \overline{OASPL}$ between 4 and 5~dBA is expected in this region.

For case L in Fig.~\ref{f5: oaspl_compar_line}~(c), the curves for $y=1480~$m and $y=1960~$m, which both correspond to the location of the turbines, match almost perfectly. The $\rm \overline{OASPL}$ increase induced by the wake effect is similar to that of the baseline case. Overall, levels are higher (about 1~dBA) downwind due to the contributions of both turbines. Finally, for case S in Fig.~\ref{f5: oaspl_compar_line}~(d), the peaks due to the wake also coincide with those of case B but are shifted by 480~m for the line at $y=1960~$m (aligned with turbine S2) because of the turbine position offset. For the line at $y=1480~$m (aligned with S1), the $\rm \overline{OASPL}$ shows for $x>1480$~m an overall increase by 3~dBA compared to case B, due to the contribution of S2. In particular, the presence of S2 mitigates the increase in $\rm \overline{OASPL}$ due to wake focusing, causing a reduction in the peak amplitude.  Upwind, the shadow zone is consistent across all cases.

\begin{figure}[htbp!]
	\centering
	\includegraphics{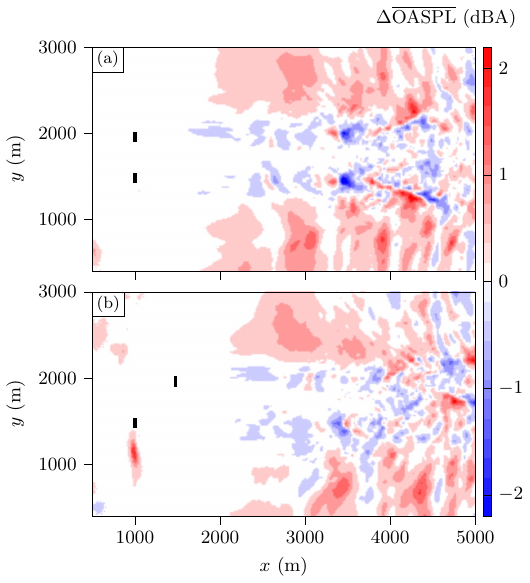}
	\caption{Difference between the actual $\overline{\rm OASPL}$ and that obtained using $\Delta L$ of turbine B  in the plane $z=2~$m  for cases (a)~L and (b)~S.}
	\label{f5: oaspl_diff_top}
\end{figure}

\begin{figure*}[h!tbp]
	\centering
	\includegraphics{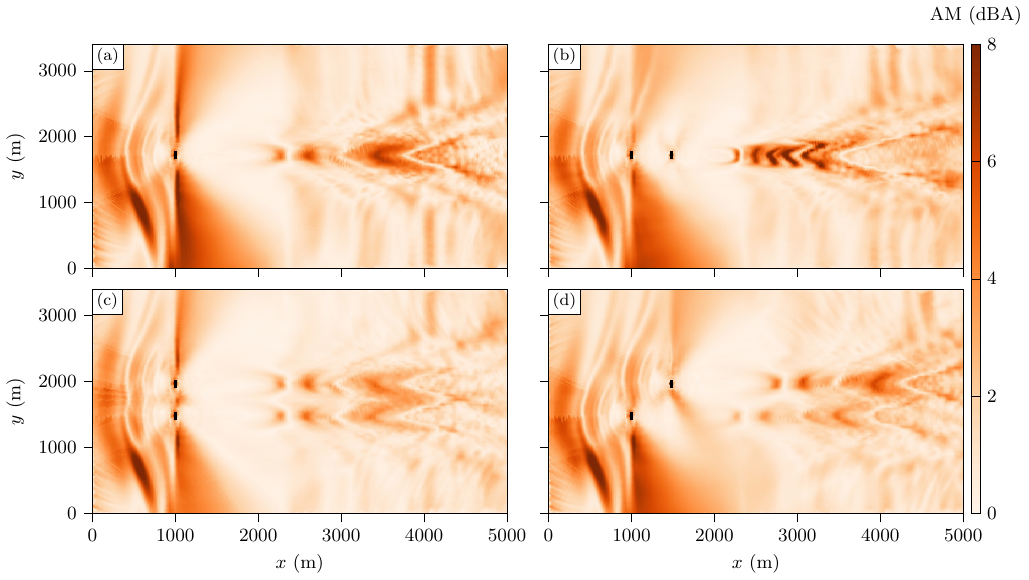}
	\caption{Amplitude modulation at $z=2~$m for the four cases: (a) B, (b) C, (c) L, and (d) S.}
	\label{f5: am_top}
\end{figure*}

To investigate the importance of the second wake on noise prediction, additional simulations are performed for cases S and L using $\Delta L$ computed for turbine B instead of the actual $\Delta L$  and keeping the same source properties. Fig.~\ref{f5: oaspl_diff_top} shows the difference in $\overline{\rm OASPL}$ between those obtained for the actual cases and those for the artificial cases in a plane at 2~m height. For case L (Fig.~\ref{f5: oaspl_diff_top}~(a)), there is an increase of 1-2~dBA induced by the second wake focusing off the $x$-axis for $x>2000$~m and for  $y<1200~$m and $y>2200~$m. In the downwind direction,  SPL decreases by less than 1~dBA for 2000~m$<x<$3200~m. At $x=3200$~m, the levels abruptly increase before decreasing. This is due to a shift in focusing position between the reference case and the artificial one. This shift may come from the aerodynamic interactions between the two turbines or from small differences in the simulated flows between cases B and L.

In case S (Fig.~\ref{f5: oaspl_diff_top}~(b)) the difference induced by the second wake in the oblique direction ($x>2000~$m and $y>2200~$m, and $x>3000~$m and $y<1200~$m) is of the same order of magnitude, despite the stronger increase in $\Delta L$ described in Sec~\ref{s5: deltaL}. In the downwind direction, a reduction by less than 1~dBA is observed for $x<4500$~m. Abrupt variations in $\rm \overline{OASPL}$ are also noticed due to a shift in the focusing patterns at the ground as for case L.

\subsection{\modif{Amplitude modulation}}

\begin{figure}[htbp!]
	\centering
	\includegraphics{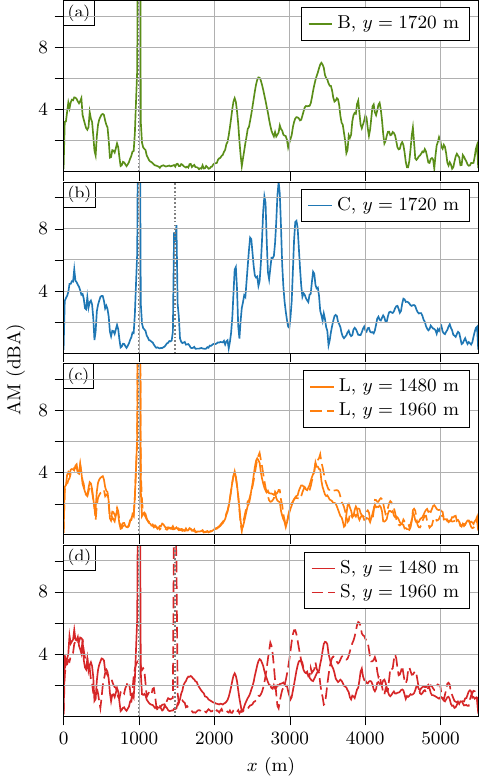}
	\caption{Amplitude modulation at $z=2~$m along lines of receivers whose $y$-coordinate corresponds to the turbine positions  for the four cases: (a) B, (b) C, (c) L, and (d) S.}
	\label{f5: am_compar_line}
\end{figure}

The AM fields at 2~m height are plotted for the four cases in Fig.~\ref{f5: am_top}. For case L, since the wind turbines operate at identical rotational speeds, AM depends on the angular offset between the two rotors. In this section, we assume no angular offset; its impact on AM is discussed in Sec.~\ref{synchr.sec}.
Case B in Fig.~\ref{f5: am_top}~(a) shows cross-wind AM induced by the blade rotation, upwind AM at the shadow zone boundary, and downwind AM due to wake focusing for $x>2000~$m. Similar AM patterns are observed for the other three cases. For case C in Fig.~\ref{f5: am_top}~(b), the AM downwind is significantly enhanced, which is expected due to the stronger focusing effect caused by the second wake. The downwind turbine C2 does not generate crosswind AM, as its acoustic power is considerably smaller than that of the upwind turbine C1. For cases L and S in Figs.~\ref{f5: am_top}~(c) and (d), both crosswind and downwind AM are lower than that in the baseline case.  This is due to the combined contributions of the two turbines, which tend to reduce the difference between minimum and maximum OASPL observed at a given receiver. For case S in Fig.~\ref{f5: am_top}~(d), downwind AM is stronger at $y=1960$~m than at $y=1480~$m. Indeed, at $y=1960$~m, the contribution of turbine S1 is small compared to that of S2 and hence the AM due to S2 is close to the one obtained for the baseline case downwind.  However, at $y=1480~$m, the contribution of S2 is significant so that AM due to S1 is reduced compared to case B.

As in Sec.~\ref{oaspl.sec}, AM is plotted along lines of receivers whose $y$-coordinate corresponds to the turbine positions at a height of 2~m for the different cases in Fig.~\ref{f5: am_compar_line}. For case B (Fig.~\ref{f5: am_compar_line}~(a)), significant values of AM  are obtained  upwind (around 4~dBA)  and downwind (around 7~dBA). For case C in Fig.~\ref{f5: am_compar_line}~(b), the downwind increase is more pronounced with AM  reaching up to 10~dBA. It also appears more localized within the range $2100~$m$<x<3500~$m, corresponding to the zone of increased $\rm \overline{OASPL}$ due to wake effects (see Fig.~\ref{f5: oaspl_compar_line}~(b)). Cases L and S in Figs.~\ref{f5: am_compar_line}~(c) and (d) exhibit similar AM patterns to case B but the maximum values are lower downwind (around 5~dBA).

\subsection{\modif{Synchronization and beating}} \label{synchr.sec}

\begin{figure}[htbp!]
	\centering
	\includegraphics{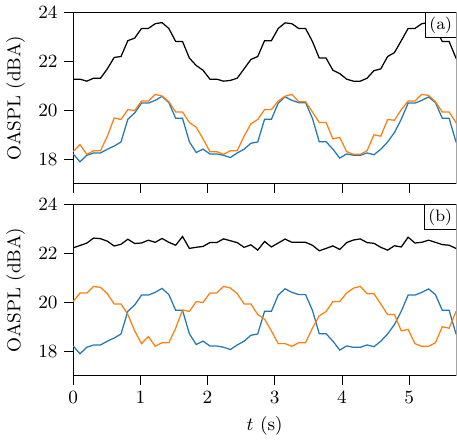}
	\caption{Temporal signals of OASPL for case L at $\mathbf{x}=(3000,1720,2)$~m for (a) $\Delta \beta_{1-2}=0$° and (b) $\Delta \beta_{1-2}=60$°. The total signal is plotted in black and the contributions of each turbine are plotted in blue and orange.}
	\label{f5: oaspl_signal_offset}
\end{figure}

For case~L, the two turbines operate under almost identical mean flow conditions and rotate at the same speed, as indicated in Tab.~\ref{t5: source}.
As a consequence, the sound field depends on the \modif{initial} relative angular position of the two turbine rotors. We denote by $\Delta \beta_{1-2}$ the offset between the angular position of the  turbine rotors (see Fig.~\ref{f5: case_def}). Fig.~\ref{f5: oaspl_signal_offset} illustrates this effect at a receiver located downwind in between the two turbines at a distance of 2~km.  In Fig.~\ref{f5: oaspl_signal_offset}~(a), the angular position of the rotor is the same for the two turbines ($\Delta \beta_{1-2} =0^{\circ}$). The time-varying SPL due to each wind turbine are in phase, resulting in AM of around 2~dBA.  In Fig.~\ref{f5: oaspl_signal_offset}~(b), the blades of the two turbines rotate with an angular offset of $\Delta \beta_{1-2} =60^{\circ}$.  In this case, the SPL signals are out of phase leading to AM close to zero. Note that this effect is not due to constructive or destructive interference. The SPL signals due to the two turbines are summed incoherently, meaning the average OASPL remains the same in both cases, despite the difference in AM.

\begin{figure*}[htbp!]
	\centering
	\includegraphics{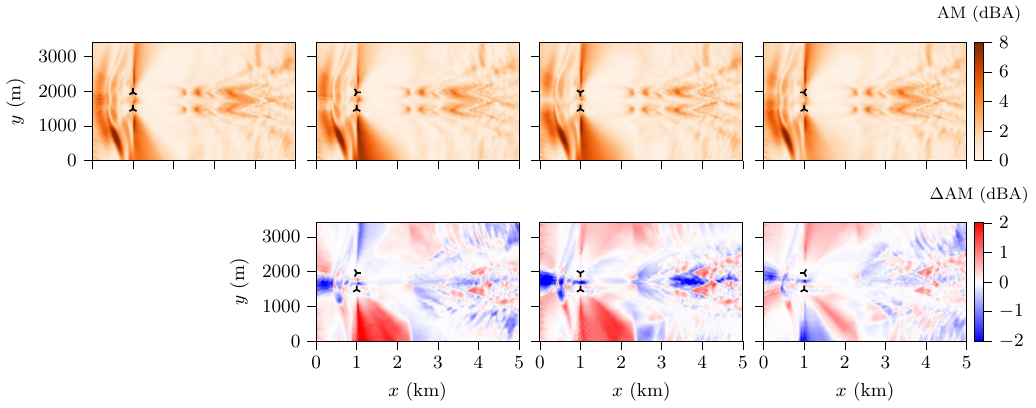}
	\caption{(top) AM in case L for several angular offsets ($\Delta \beta_{1-2}$) between the turbine rotors. From left to right, $\Delta \beta_{1-2}=0$, 30, 60, and 90°. (bottom) Difference between AM obtained with several offsets  and AM without offset ($\Delta \beta_{1-2}=0$°). Turbine positions are represented by the three blades of the rotor to help visualize the angular offset.}
	\label{f5: am_diff_top}
\end{figure*}

We further illustrate this effect in Fig.~\ref{f5: am_diff_top}. The AM fields at 2~m height are presented for several offsets between the turbine rotors. The difference between the AM obtained  for a given value of $\Delta \beta_{1-2}$ and that for $\Delta \beta_{1-2}=0$° is also depicted.
The AM patterns remain qualitatively similar for the different angular offsets, consistently exhibiting increased AM in the crosswind, upwind, and downwind directions. However, AM magnitude depends on  $\Delta \beta_{1-2}$. In the zone exactly between both turbines downwind (around $x=3500~$m) and upwind (around $x=500$~m), AM decreases as $\Delta \beta_{1-2}$ increases and reaches almost 0~dBA for $\Delta \beta_{1-2}=60$°, which corresponds to the turbines being in phase opposition. On the contrary, the crosswind AM increases by up to 2~dBA when the turbines are out of phase.

Whether the signals from the two turbines are in or out of phase depends on several factors, including the angular offset between the turbine rotors and the propagation time.
Finally, we remind the reader that if the rotational speeds of the turbines are different, the \modif{time averaged} properties of the sound field do not depend on the \modif{initial} relative angular position of the turbine rotors \modif{as the angular offset between the two rotors is not constant and varies with time}.

In cases C and S, the turbine rotors do not rotate at the same speed.
Consequently, the temporal signal of OASPL due to each turbine has a different period.
When the rotational speeds are close, as in case S, a beating effect can be observed at positions where the contributions of both turbines are of the same order of magnitude.
In these locations, the signal is modulated by an envelope with a frequency lower than the blade‑passing frequency.
\modif{The beating effect was reported in field experiments for stable atmospheric conditions in  \citet{vandenbergBeatGettingStronger2005}. This phenomenon introduces a second layer of variability with intermittency in AM. From a perceptual point of view, this suggests  that AM can be likely imperceptible to occasionally perceptible~\citep{nguyenLongtermQuantificationCharacterisation2021}.  
}

\begin{figure}[htbp!]
	\centering
	\includegraphics{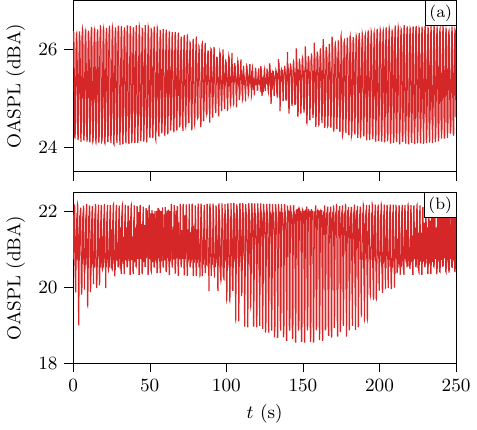}
	\caption{Temporal signals of OASPL for case S at (a) $\mathbf{x}=(1500,2500,2)~$m and (b) $\mathbf{x}=(4400,1720,2)~$m.}
	\label{f5: oaspl_signals_S2}
\end{figure}

As an example, the OASPL signals for case S at two receiver locations are plotted in Fig.~\ref{f5: oaspl_signals_S2}. Fig.~\ref{f5: oaspl_signals_S2}~(a) is obtained for a receiver located crosswind and Fig.~\ref{f5: oaspl_signals_S2}~(b) for a downwind receiver located between the two turbines.
The signals at both positions alternate between discernible AM (2~dBA crosswind and 3~dBA downwind) and much smaller AM (close to zero crosswind and 1~dBA downwind).

The carrier frequency $f_c$ and envelope frequency $f_e$ of the OASPL signal can be derived from trigonometric relations, yielding:
\begin{equation} \label{beating.eq}
	f_c=\frac{f_1+f_2}{2},\quad f_e=\frac{f_1-f_2}{2},
\end{equation}
where $f_1$ and $f_2$ are the blade passing frequencies of each turbine \modif{($f=3\Omega/(2\pi))$}.
Consequently, the carrier and envelope frequencies are identical at different receiver positions. Considering the rotational speeds of the turbines S1 and S2 in Tab.~\ref{t5: source}, \modif{we obtain a carrier frequency $f_c=0.5125~$Hz, which is close to the blade passing frequencies of the two turbines. The value of the envelope frequency $f_e=0.0025~$Hz corresponds to a period of 400~s, which matches that observed for the signals in Fig.~\ref{f5: oaspl_signals_S2} (only a half period of 200~s is visible for the envelope)}.
Moreover, Fig.~\ref{f5: oaspl_signals_S2} shows that the largest values of AM occur at different times between crosswind and downwind receivers due to varying propagation times between the sources and the receiver.
\modif{The strong beating effect is also due to a consistent rotational speed for the entire time of the simulation.
	In real life conditions, rotational speed would vary according to the wind speed and the wind turbine controller leading to a variation in the beating pattern.
}

\modif{
	In case C, the envelope and carrier frequencies are equal to $f_c=0.4325~$Hz and $f_e=0.0775$Hz.
	Here, the difference between the carrier and envelope frequencies is smaller leading to no distinct envelope modulating the signals.
	Hence, the corresponding plots are omitted for brevity.
}

\section{Conclusion} \label{conclu.sec}

\modif{
	A numerical study of noise propagation from a pair of wind turbines was presented, with a focus on how the wake effects of two turbines influence the acoustic field based on their relative positioning.
}
The most significant effects occur when the turbines are aligned with the flow direction. For the upstream turbine, the second wake enhances wake focusing leading to increased average OASPL and AM downwind compared to an isolated wind turbine. The downstream turbine exhibits a significant reduction in noise emission due to decreased inflow velocity at the rotor plane, resulting from its position within the wake of the upstream turbine.

When the two turbines are side-by-side or staggered,
\modif{the influence of the second wake on noise propagation is moderate}.
An increase in the average OASPL by less than 2~dBA has been noticed in the oblique direction downwind. Moreover, AM is reduced compared to that of an isolated wind turbine, since the superposition of signals from both turbines tends to smooth out the noise level fluctuations, which reduces the difference between maximum and minimum sound pressure levels.

An important aspect of the analysis involves the temporal evolution of the OASPL at a receiver. When turbines rotate at identical speeds, the angular offset between the rotors significantly impacts the temporal signal, leading to variations in AM. However, when rotor speeds differ slightly, the signal exhibits prominent beating effects, creating clear intermittency in AM. This behavior is especially important because it highlights a strong sensitivity of AM characteristics to small differences in rotor dynamics. Understanding this sensitivity is crucial for accurately capturing noise variability and its perceptual effects.

Future work could include comparison with experiments and a relaxation of current modeling assumptions, particularly by accounting for three-dimensional propagation and scattering by turbulence. Effects of wind turbine interactions on noise propagation could also be investigated  at wind farm level for further understanding. Finally, integration of these interaction effects into operational models could be studied.

\begin{acknowledgments}
	This work was performed within the framework of the LABEX CeLyA (ANR-10-LABX-0060) of Universit\'e de Lyon, within the program ``Investissements d’Avenir" (ANR-16-IDEX-0005) operated by the French National Research Agency (ANR). This project has received funding from the European Research Council under the European Union's Horizon Europe program (Grant No. 101124815). It was granted access to the HPC resources of PMCS2I (P\^ole de Mod\'elisation et de Calcul en Sciences de l'Ing\'enieur et de l'Information) of Ecole Centrale de Lyon.  This work was supported by the Franco-Dutch Hubert Curien partnership (Van Gogh Programme No.~49310UM). For the purpose of Open Access, a CC-BY public copyright license has been applied by the authors to the present document and will be applied to all subsequent versions up to the Author Accepted Manuscript arising from this submission.
\end{acknowledgments}

\section*{Author Declarations}

The authors have no conflicts to disclose.

\section*{Data Availability}

Data available on request from the authors.

\end{document}